\documentstyle[epsf,12pt]{article}
\begin{document}
\setlength\unitlength{1pt} 

\begin{center}

{\Large\bf Pole masses of quarks \\ ~ \\ in dimensional reduction}

\vfill

{\Large L.~V. ~Avdeev
\footnote{~E-mail: $avdeevL@thsun1.jinr.dubna.su$}}
and
{\Large M.~Yu. ~Kalmykov
\footnote{~E-mail:$kalmykov@thsun1.jinr.dubna.su$}}

{\em Bogoliubov Laboratory of Theoretical Physics, Joint Institute for
Nuclear Research, RU-$141980$~Dubna (Moscow Region), Russian Federation}

\vfill

\begin{abstract}

Pole masses of quarks in the quantum chromodynamics are calculated to
the two-loop order in the framework of the regularization by
dimensional reduction. For the diagram with a light quark loop, the
non-Euclidean asymptotic expansion is constructed with the external
momentum on the mass shell of a heavy quark.

\end{abstract}
\end{center}

\vfill

\thispagestyle{empty}
\setcounter{page}0
\newpage

\section{Introduction}

Regularization by dimensional reduction, a modification of the
conventional dimensional regularization, was proposed \cite{Siegel}
as a candidate for an invariant regularization in supersymmetric
theories.  The idea is very simple. Supersymmetry is only possible in
certain integer dimensions (four, to be specific). Therefore, the
vector and spinor algebra in the numerator of Feynman diagrams should
be retained four-dimensional.  On the other hand, regularization of
momentum integrals is achieved in $N=4-2\varepsilon$ dimensions. The
possibility of canceling the squared momenta in the numerator and the
denominator, crucial for maintaining the gauge invariance, requires
then that the momenta form an $N$-dimensional subspace in four
dimensions.

Such a dimensional reduction from a finite-dimensional space into the
formal space of the regularization proves to be mathematically
inconsistent \cite{inconsistency}. The regularized space should be
recognized as infinite-dimensional because antisymmetrization over an
arbitrary number of indices does not give the identical zero.  A
consistent modification \cite{AV} disables any strictly
four-dimensional objects, like the totally antisymmetric $\epsilon$
tensor, thus distorting the algebra above a certain order.  On the
other hand, the original version, in spite of being globally
inconsistent, can provide unambiguous results until
antisymmetrization over five indices actually comes into play
(in particular, the evaluation of the quark propagator in the 
quantum chromodynamics may became ambiguous not earlier than 
at the four-loop level). 
Besides, it is very convenient technically, to have the usual
four-dimensional algebra of spinor matrices (including $\gamma_5$)
and even their Fierz rearrangement \cite{Fierz}. This is why the
regularization by dimensional reduction was very soon applied to
nonsupersymmetric calculations \cite{Italy} in the Standard Model. It
was also used, to verify the consistency of the $\gamma_5$
prescription in a three-loop QCD-electroweak calculation \cite{rho}.

In the present paper we demonstrate the two-loop calculation of the
pole masses of quarks in quantum chromodynamics in the framework of
the regularization by dimensional reduction. The corresponding
calculation in the conventional dimensional regularization was done
in refs.~\cite{else,broad}. The leading approximation of the present
calculation in dimensional reduction (that ignored contributions of
heavier quarks and masses of lighter quarks) was used in
ref.~\cite{rho}, to verify the $\gamma_5$ prescription.

Here we evaluate the contributions of the heavier and lighter quarks
by means of the asymptotic expansion, which is of a special
methodical interest. While the structure of the large-mass expansion
caused no particular doubts, being defined by the universal Euclidean
rules \cite{rules}, the expansion in a small mass is more
complicated. Its definition needs an on-shell infrared extension
which we construct explicitly for the present case.

\section{The choice of the renormalization scheme}

Let us discuss the dimensional reduction in some more detail. As the
momenta become $N$-dimensional, 4-vectors naturally split into true
$N$-vectors and so-called $\varepsilon$ scalars which fall into a
complementary subspace of dimension $4-N=2\varepsilon$, orthogonal to
the momenta. The  $\varepsilon$ scalars are nothing else but matter
fields. Their presence is the only difference from the conventional
dimensional regularization. Order by order in perturbation theory,
contributions of $\varepsilon$ scalars are equivalent to finite
counterterms, that is, to a change of the renormalization scheme.
When performing renormalizations, one should remember an important
fact. Formal 4-covariance (valid at the stage of generating the
`bare' diagrams) may be broken by counterterms. Renormalizations of
$\varepsilon$ scalars and of their interactions (including the pole
contributions in $\varepsilon$ ) are not identical to those of the
vectors.

Moreover, quantum corrections may also generate a mass for
$\varepsilon$ scalars. There is an arbitrariness in choosing the
renormalization scheme for this mass \cite{jones}. A consistent way
is to choose the finite $\varepsilon$-scalar mass counterterm so that
the pole (and renormalized) mass of the $\varepsilon$ scalars be
zero.  The $\varepsilon$-scalar field renormalization is left
minimal. Then no additional dimensional parameter is ever involved in
renormalizations while the polynomial renormalization of other
masses, independence of renormalizations of dimensionless coupling
constants on masses, and gauge independence of all $\beta$ functions
are retained. We follow that scheme in the present calculation.
However, this may not be the only way. For example, in a softly
broken supersymmetric Yang--Mills theory it would be natural trying
to restore the maximum symmetry by relating the mass of $\varepsilon$
scalars to the gluino mass as a special solution to the
renormalization-group equations in the minimal subtraction scheme.

One should also bear in mind technical complications that are
inevitable when performing a massive calculation up to a finite part
with an additional mass on previously massless lines.

Other renormalizations are done minimally, that is by subtracting
only poles in $\varepsilon$. We use a modified $\overline{\rm MS}$
definition \cite{ms,broad}, dividing each loop by $(4\pi)^\varepsilon
\Gamma(1+\varepsilon)$ rather than multiplying by
$\Gamma(1-\varepsilon)$. This definition is more convenient in
calculations with masses because a simple massive loop is then a pure
rational function of $N$ without any $\Gamma$ functions. It would be
proper to call the scheme MMS (massive minimal subtractions).

\section{The pole mass}

The singularity mass of a particle is a physically meaningful  
quantity \cite{else}.  We restrict ourselves to perturbation theory 
only and do not analyze the exact nature of the two-point function 
singularity which may involve a branching point. In the present paper 
we accept the conventional term  'pole mass'  which  refers to the 
point where the real part of the inverse propagator turns into zero. 

Being a physical quantity, the pole mass is renormalization- and 
gauge-invariant. We verify the gauge invariance
by performing our calculations in the arbitrary covariant gauge.

The pole mass of a quark $m_P$ is defined as a formal solution for
$\hat p$ (in the Minkowski metric) at which the reciprocal of the
connected full propagator equals zero:

\begin{equation}
\hat p - m - \Sigma(\hat p, m) = 0,
\label{pole}
\end{equation}

\noindent
where $\Sigma(\hat p, m) = \hat p ~ A(p^2,m) + m ~ B(p^2,m)$ is the
one-particle-irreducible two-point function (including the $i$ factor
for one of its legs); $m$ may stand for the bare or renormalized mass,
$m_B$ or $m_R$, depending on the prescription used in evaluating
$\Sigma$. The solution to eq.~(\ref{pole}) is sought order by order in
perturbation theory. To two loops

\begin{equation}
m_P = m +\Sigma_1(m,m) +\Sigma_2(m,m)
+ \Sigma_1(m,m)~ \Sigma_1'(m,m) + {\cal O}(\Sigma_3),
\label{2loop}
\end{equation}

\noindent where $\Sigma_L$ is the $L$-loop contribution to $\Sigma$,
and the prime denotes the derivative with respect to the first argument.

According to eq.~(\ref{2loop}), technically, we need to evaluate
propagator-type diagrams up to two loop on shell.  For the one-loop
diagrams, also the derivative with respect to the momentum is
necessary. It can be expressed through the derivative with respect to
the mass because the total dimension of the diagrams is known.  Most
of these calculations can be performed with the aid of the SHELL2
package \cite{shell2} intended to analytically compute 
propagator-type Feynman integrals that involve a continuous line of 
one mass and the external momentum on the mass shell of that mass 
up to two loops.  We use a modified implementation of the
package, advanced in the scope and efficiency.  The original
algorithm \cite{shell2} for reducing scalar numerators is only
applied to the first powers of the numerators. In other cases we use
the strategy described in ref.~\cite{proto} of employing a recurrence
relation for a neighbor triangle.

\begin{figure}[htbp]
$$
\begin{picture}(150,50)(-75,-25)
\put(-20,0){\oval(8,8)[lt]}
\put(-12,8){\oval(8,8)[lt]}
\put(-20,8){\oval(8,8)[br]}
\put(20,0){\oval(8,8)[rt]}
\put(12,8){\oval(8,8)[rt]}
\put(20,8){\oval(8,8)[bl]}
\put(24,1){\line(1,0){10}}
\put(24,-1){\line(1,0){10}}
\put(-24,1){\line(-1,0){10}}
\put(-24,-1){\line(-1,0){10}}
\put(0,12){\circle{24}}
\put(0,0){\oval(48,48)[b]}
\put(0,0){\oval(44,44)[b]}
\put(-23,0){\circle*3}
\put(23,0){\circle*3}
\put(-12,12){\circle*3}
\put(12,12){\circle*3}
\put(24,-25){(a)}
\end{picture}
\begin{picture}(150,50)(-75,-25)
\put(-20,0){\line(1,1){20}}
\put(16,0){\oval(8,8)[rt]}
\put(8,8){\oval(8,8)[rt]}
\put(0,16){\oval(8,8)[rt]}
\put(16,8){\oval(8,8)[bl]}
\put(8,16){\oval(8,8)[bl]}
\put(-16,-4){\circle*1}
\put(-12,-8){\circle*1}
\put(-8,-12){\circle*1}
\put(-4,-16){\circle*1}
\put(-0.7,-19.3){\line(1,1){20}}
\put(0.7,-20.7){\line(1,1){20}}
\put(20,1){\line(1,0){10}}
\put(20,-1){\line(1,0){10}}
\put(-20,1){\line(-1,0){10}}
\put(-20,-1){\line(-1,0){10}}
\put(0,20){\line(0,-1){40}}
\put(-20,0){\circle*3}
\put(20,0){\circle*3}
\put(0,20){\circle*3}
\put(0,-20){\circle*3}
\put(-18,10){1}
\put(12,10){2}
\put(-15,-22){3}
\put(12,-22){4}
\put(2,-5){5}
\put(25,-25){(b)}
\end{picture}
$$
\caption{ \label{loop}
The diagram with a quark loop (a) and its scalar prototype
(b).}
\end{figure}
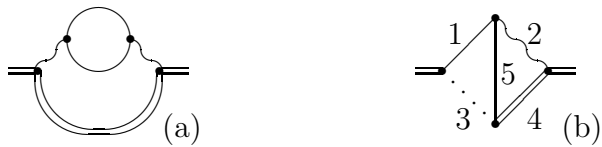

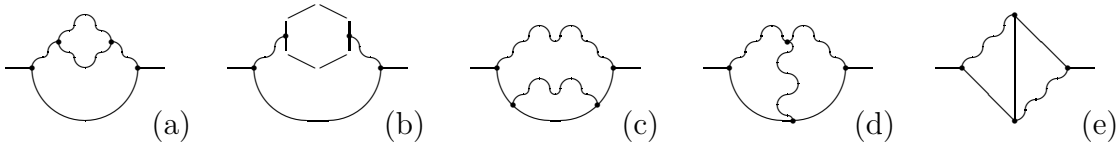
\begin{figure}[bthp]
\centerline{
\begin{picture}(80,75)(-40,-25) 
\put(-30,0){\line(1,0){10}}
\put(0,0){\oval(40,40)[b]}
\put(20,0){\line(1,0){10}}
\put(-16,0){\oval(8,8)[lt]}
\put(-16,6){\oval(4,4)[br]}
\put(-10,6){\oval(8,8)[lt]}
\put(-6,10){\oval(8,8)[l]}
\put(-6,16){\oval(4,4)[br]}
\put(0,16){\oval(8,8)[t]}
\put(6,16){\oval(4,4)[bl]}
\put(6,10){\oval(8,8)[r]}
\put(6,4){\oval(4,4)[lt]}
\put(0,4){\oval(8,8)[b]}
\put(-6,4){\oval(4,4)[tr]}
\put(10,6){\oval(8,8)[tr]}
\put(16,6){\oval(4,4)[bl]}
\put(16,0){\oval(8,8)[tr]}
\put(-20,0){\circle*2}
\put(-10,10){\circle*2}
\put(10,10){\circle*2}
\put(20,0){\circle*2}
\put(25,-25){(a)}
\end{picture}
\begin{picture}(88,75)(-44,-25) 
\put(-24,0){\line(-1,0){10}}
\put(0,0){\oval(48,40)[b]}
\put(24,0){\line(1,0){10}}
\put(-20,0){\oval(8,8)[lt]}
\put(-20,8){\oval(8,8)[br]}
\put(-12,8){\oval(8,8)[lt]}
\put(-12,6.5){\line(0,1){11}}
\put(-11,19){\line(2,1){10}}
\put(1,24){\line(2,-1){10}}
\put(12,17.5){\line(0,-1){11}}
\put(11,5){\line(-2,-1){10}}
\put(-1,0){\line(-2,1){10}}
\put(12,8){\oval(8,8)[tr]}
\put(20,8){\oval(8,8)[bl]}
\put(20,0){\oval(8,8)[tr]}
\put(-24,0){\circle*2}
\put(-12,12){\circle*2}
\put(12,12){\circle*2}
\put(24,0){\circle*2}
\put(25,-25){(b)}
\end{picture}
\begin{picture}(84,50)(-42,-25) 
\put(-32,0){\line(1,0){10}}
\put(0,0){\oval(44,40)[b]}
\put(22,0){\line(1,0){10}}
\put(-18,0){\oval(8,8)[lt]}
\put(-18,6){\oval(4,4)[br]}
\put(-12,6){\oval(8,8)[lt]}
\put(-12,12){\oval(4,4)[br]}
\put(-6,12){\oval(8,8)[t]}
\put(0,12){\oval(4,4)[b]}
\put(6,12){\oval(8,8)[t]}
\put(12,12){\oval(4,4)[bl]}
\put(12,6){\oval(8,8)[tr]}
\put(18,6){\oval(4,4)[bl]}
\put(18,0){\oval(8,8)[tr]}
\put(-12,-14){\oval(8,8)[lt]}
\put(-12,-8){\oval(4,4)[br]}
\put(-6,-8){\oval(8,8)[t]}
\put(0,-8){\oval(4,4)[b]}
\put(6,-8){\oval(8,8)[t]}
\put(12,-8){\oval(4,4)[bl]}
\put(12,-14){\oval(8,8)[tr]}
\put(-22,0){\circle*2}
\put(22,0){\circle*2}
\put(-16,-14){\circle*2}
\put(16,-14){\circle*2}
\put(25,-25){(c)}
\end{picture}
\begin{picture}(84,50)(-42,-25) 
\put(-32,0){\line(1,0){10}}
\put(0,0){\oval(44,40)[b]}
\put(22,0){\line(1,0){10}}
\put(-18,0){\oval(8,8)[lt]}
\put(-18,6){\oval(4,4)[br]}
\put(-12,6){\oval(8,8)[lt]}
\put(-12,12){\oval(4,4)[br]}
\put(-6,12){\oval(8,8)[t]}
\put(0,12){\oval(4,4)[b]}
\put(6,12){\oval(8,8)[t]}
\put(12,12){\oval(4,4)[bl]}
\put(12,6){\oval(8,8)[tr]}
\put(18,6){\oval(4,4)[bl]}
\put(18,0){\oval(8,8)[tr]}
\put(0,8){\oval(4,4)[r]}
\put(0,2){\oval(8,8)[l]}
\put(0,-6){\oval(8,8)[r]}
\put(0,-14){\oval(8,8)[l]}
\put(0,-20){\oval(4,4)[tr]}
\put(-22,0){\circle*2}
\put(22,0){\circle*2}
\put(2,-20){\circle*2}
\put(0,10){\circle*2}
\put(25,-25){(d)}
\end{picture}
\begin{picture}(80,50)(-40,-25) 
\put(-30,0){\line(1,0){10}}
\put(-20,0){\line(1,-1){20}}
\put(0,-20){\line(0,1){40}}
\put(0,20){\line(1,-1){20}}
\put(20,0){\line(1,0){10}}
\put(-16,0){\oval(8,8)[lt]}
\put(-16,8){\oval(8,8)[br]}
\put(-8,8){\oval(8,8)[lt]}
\put(-8,16){\oval(8,8)[br]}
\put(0,16){\oval(8,8)[lt]}
\put(0,-16){\oval(8,8)[br]}
\put(8,-16){\oval(8,8)[lt]}
\put(8,-8){\oval(8,8)[br]}
\put(16,-8){\oval(8,8)[lt]}
\put(16,0){\oval(8,8)[br]}
\put(-20,0){\circle*2}
\put(0,-20){\circle*2}
\put(0,20){\circle*2}
\put(20,0){\circle*2}
\put(25,-25){(e)}
\end{picture}
}
\caption{Further two-loop diagrams, contributing to the pole mass of a
quark, which involve no other nonzero masses. Solid lines correspond to
quarks, wavy lines to gluons, and dashed lines to the Faddeev--Popov
ghosts.}
\label{others}
\end{figure}

At the two-loop level there are 6 diagrams contributing to the pole
mass of a quark. They are presented in fig.~\ref{loop}(a) and
fig.~\ref{others}.

In individual diagrams there may be on-shell infrared divergences.
How\-ever, the complete pole mass is infrared-finite ~\cite{else}. It
only involves ultraviolet poles in $\varepsilon$, which are removed
after introducing ultraviolet counterterms. The counterterm can be
evaluated in some other infrared-secure off-shell configuration of
momenta and masses. The renormalization of the Yukawa charge for
$\varepsilon$ scalars and their (nonminimal) mass counterterm are
calculated separately. The point is that in a nonsupersymmetric
theory, like QCD, the (ultraviolet divergent) renormalization
constants for the Green functions of $\varepsilon$ scalars are quite
different from those of the vectors (the latter being restricted by
gauge invariance).  Renormalizing a physical quantity, as the pole
mass, we can accumulate the difference in the bare charge
counterterms alone, ignoring the field renormalizations, if we
evaluate the contributions of counterterms for the whole sum of the
diagrams.

\section{The rules of asymptotic expansions}

Let us concentrate on the diagram with a quark loop
[fig.~\ref{loop}(a)].  When the mass of the quark in the loop $m_1$
is different from the mass of the external quark $m$, this diagram
cannot be calculated by the SHELL2 algorithms. In the conventional
dimensional regularization, the result involves complicated
transcendental functions \cite{broad}. We are going to demonstrate
how to evaluate the diagram by means of expanding it in the ratio of
the masses. If the mass in the loop is large as compared to the mass
of the propagating quark, $m_1 \gg m$, then the structure of the
asymptotic expansion \cite{rules} established in case of purely
Euclidean expansions does not require any modifications.  The rules
are as follows. The expansion is a sum over 'ultraviolet' subgraphs
of the diagram. An ultraviolet subgraph must contain all lines with
large masses, the points were large external momenta (if any) flow
in/or out. The large momenta ought to go only through the ultraviolet
subgraph and obey the momentum conservation law.  And last, an
ultraviolet subgraph should be one-particle irreducible with respect
to lines with small and zero masses although may consist of several
disconnected parts. An ultraviolet subgraph is Taylor-expanded in its
small parameters (external momenta and internal masses) and then
shrunk to a point and inserted in the numerator of the remaining
Feynman integral.

The large-mass expansion of the diagram, fig.~\ref{loop}(a), involves
only two ultraviolet subgraphs: the whole graph and the loop of two
lines with the heavy mass.  As result of the expansion two-loop
bubble integrals and propagator-type integrals on shell arise.

The case $m_1\ll m$ is more difficult. Naively applying the standard
expansion, we would encounter actual infinity for the subgraph that
includes only the internal `on-shell' line with the mass $m$. Thus, the
expansion of this ultraviolet subgraph needs some extended definition.

\section{The scalar prototype}

It is convenient to perform further considerations in terms of the
scalar prototype of the diagram [fig.~\ref{loop}(b)] in Euclidean
space (after the Wick rotation). The prototype involves arbitrary
integer powers of the scalar denominators $c_L = k_L^2 + m_L^2$ on
the lines.  Their powers $j_L$ are called indices of the lines. The
mass-shell condition for the external momentum now is $p^2=-m^2$. Any
scalar products of the momenta in the numerator are reduced to
changed powers of the scalar invariants in the denominator. Thus, the
indices may sometimes become negative. In particular, an auxiliary
line, of the prototype, line \#~3, is generated with an always
non-positive index $j_3\le 0$.  From the viewpoint of the $R$
operation or asymptotic expansion, line~3 represents just a vertex
(local in co-ordinate space). It is convenient to choose
$m_3^2 = m^2 + m_1^2$, so that $p k_1 = \frac{1}{2}(c_1-c_3)$.
Other masses are $m_2=0$, $m_4=m$, $m_5=m_1$.

The index of the auxiliary line $j_3$ can always be reduced to zero by
means of recurrence relations. The recurrence relations are derived by
the Chetyrkin--Tkachov method of integration by parts \cite{parts}. We
use a shorthand notation $\{XYZ\}$ of ref.~\cite{proto} to denote the
relation for the triangle formed of lines \# $X$, $Y$, and $Z$
(the latter is optional, a degenerate case being allowed):
\newcommand\eqnum[1] {\eqno{#1}}

$$ \int \frac{d^N k_X}{c_X^{j_X} c_Y^{j_Y} c_Z^{j_Z}}
\Big( N -2 j_X -j_Y -j_Z +j_X \frac{2 m_X^2}{c_X}
+j_Y\frac{m_X^2+m_Y^2-m_{XY}^2+c_{XY}-c_X}{c_Y}
$$
$$
+j_Z\frac{m_X^2+m_Z^2-m_{XZ}^2+c_{XZ}-c_X}{c_Z} \Big) = 0 ,
\eqnum{\{XYZ\}} $$

\noindent where a double index $XY$ refers to the line that starts at
the point where lines $X$ and $Y$ meet. For an external line on the mass
shell, the value of $c_L$ is equal to zero. Expressing some term of
eq.~$\{XYZ\}$ through other terms, we obtain a relation between Feynman
integrals of the same prototype but with varied indices of the lines.

In order to eliminate a numerator $j_3<0$, we apply the following
relations. If $j_5 \ne 1$, we solve equation eq.~\{425\} with respect
to the $c_3/c_5$ term. Other wise, if $j_4 \ne 1$, the $c_3/c_4$
term of eq.~\{524\} is used. If $j_1 \ne 1$, the $c_3$ numerator is
reduced using eq.~\{315\} + \{425\}. In case $j_1 = j_4 = j_5 = 1$ we
apply eq.~\{135\} + \{245\}, solved with respect to the free term, to
create a denominator for which the above relations are applicable.

At  $j_3 = -1$, one can derive a simpler formula which is more efficient
than the above algorithm. First of all, $c_3 = c_1 - 2 p k_1$. On
integration over $k_1$ with just lines 1 and 5 involved, by Lorentz
covariance, a single component of $k_{1 \mu}$ turns into $k_{2 \mu}$
with the coefficient $k_1 k_2/k_2^2 $. Hence, $p k_1$ is equivalent to
$(p k_2)(k_1 k_2)/k_2^2 = \frac{1}{4}(c_2-c_4)(c_1+c_2-c_5)/c_2$. Thus,
the first power of the numerator on line 3 can be replace as
$c_3 = \frac{1}{2}[ c_1 + c_4 + c_5 + (c_1-c_5) c_4/c_2 - c_2 ]$.

\section{The small mass expansion}

We assume that the general $R$-operation-like structure of asymptotic
expansion is retained in the case of the on-shell expansion in a small
mass. The set of ultraviolet subgraphs of our diagram (after
eliminating line 3) is:
\{1, 2, 4, 5\}, \{1, 2, 4\}, \{2, 4, 5\}, \{4\}.
The first three subgraphs do not cause any technical complications
in their evaluation. The whole graph becomes a two-loop on-shell
propagator-type integral which is a particular case of a SHELL2
prototype with two massive lines, one of the latter being contracted.
The two succeeding subgraphs both yield a product of a one-loop bubble
integral by an on-shell propagator type integral. The last subgraph
\{4\} is infrared-singular, its formal Taylor-expansion being
infinite on mass shell.  Let us try to derive an extended definition
for the asymptotic expansion of this subgraph. We need to expand the
propagator $1/c_4 = 1/(-2 p k_2 + k_2^2)$ in the small momentum $k_2$.
It is reasonable to make use of the expansion

\begin{equation} 1/(-2 p k_2 + k_2^2) =
\sum_{n=0}^\infty (k_2^2)^n/(-2 p k_2 + 0)^{n+1} , \label{c4}
\end{equation}

\noindent where $+0$ keeps the correct causal (Euclidean) rule of going
around the singularity at zero which appears as we perform the on-shell
expansion.

The even powers of the scalar product in eq.~(\ref{c4}), for which $+0$ is
inessential in the context of dimensional regularization,
can be worked up by the formula

\begin{equation} \int d^N k_2 ~ (p k_2)^{2n} f(k_2^2) =
\frac{\Gamma(N/2) \Gamma(1/2+n)}{\Gamma(N/2+n) \Gamma(1/2)} \int d^N
k_2 ~ (p^2 k_2^2)^n f(k_2^2) , \label{even}
\end{equation}

\noindent where $f$ is an arbitrary function which depends on
$k_2$ through $k_2^2$ only.
The formula can be proved by induction based on the Lorenz structure
of the result, for positive integer values of $n$ and then analytically
continued to arbitrary complex values. It is regular for negative
integer $n$. We assume that $j_3$ has been brought to zero by recurrence
relations before the expansion. Then, subgraph $\{1, 5\}$ after
integration over $k_1$ depends only on $k_2^2$. This fact permits us to
apply eq.~(\ref{even}) to even negative powers of $p k_2$ in eq.~(\ref{c4}).
The integral can then be explicitly evaluated and transformed to the
form

\begin{eqnarray}
&&\int \frac{d^N k_1 ~d^N k_2}
{\pi^N \Gamma^2(3-N/2) c_1^{j_1} c_2^{j_2} c_5^{j_5} (p k_2)^{2j}}
=
\nonumber \\*
&&
\frac{ (m_1^2)^{N-j_1-j_2-j_5-j} \Gamma(1/2) \Gamma(N/2-j_2-j) }
{(-p^2)^j \Gamma(j_1) \Gamma(j_5) \Gamma(1/2+j) \Gamma(N/2-j) }
\nonumber \\*
&&
\times
\frac{\Gamma(j_1+j_2+j-N/2)
\Gamma(j_5+j_2+j-N/2)\Gamma(j_1+j_2+j_5+j-N)}
{\Gamma(j_1+j_5+2j_2+2j-N)}.
\label{int}
\end{eqnarray}

Odd negative powers reveal on-shell infrared singularities of the
propagator, which ought to generate some counterterms as a part of the
asymptotic expansion. They become manifest as we remember that

\begin{equation} 1/(p k_2 +0)^{2n+1} = 1/(p k_2)^{2n+1} +
\frac{\pi}{(2n)!} ~ \delta^{(2n)}(p k_2)
\label{odd}
\end{equation}

\noindent for purely imaginary $p$ as we have it. The first term on the
right-hand side of eq.~(\ref{odd}), which generates the integral in the
sense of the principal value, yields just zero with a function of
$k_2^2$. The second term picks out exactly the $2n$'th coefficient in
the Taylor expansion of the rest of the diagram in $p k_2$, that is
the component of $k_2$ parallel to $p$. This is the kind of infrared
counterterms, only partially local in momentum space, which are
characteristic of the non-Euclidean asymptotic expansion.  The delta
function itself takes off one integration over the corresponding
component of $k_2$, leaving the integral in dimension $N-1$.

The seemingly independent treatment of the odd powers of the scalar
product in the denominator by eq.~(\ref{odd}) is in fact firmly bound
to formula (\ref{even}) for the even powers in the general framework of
dimensional regularization. The point is that the latter always implies
an auxiliary intermediate analytic regularization in the line
indices~\footnote{ L.V. Avdeev is grateful to Prof. K.G. Chetyrkin for
a clarifying discussion on the subject in 1983.}. One seeks a
correlated region of all the parameters, the powers and the space-time
dimension, in which the integral under consideration is well defined
and analytic in the parameters. Then the resulting function is
analytically continued to the whole complex plane, and eventually, the
limit is taken in the indices to the values that we started with. As a
rule, the regular limit does exist while the space-time dimension is
kept noninteger. According to this schedule, we have evaluated the
integral on the right-hand side of eq.~(\ref{even}) for the relevant
diagram, using the power of the scalar product as an additional
regulator rather than a natural number and taking the limit
$n \to -j$, eq.~(\ref{int}). Having identically transformed the result,
we can try to take the limit for half-integer values of $j$ as well.
The limit does always exist and exactly coincides with the value that we
get from $\int d^{N-1} k_2$ according to the eq.~(\ref{odd}). It is
interesting to note that the odd coefficients of the expansion turn out
to be ultraviolet-finite, that is allow a smooth $N \to 4$ limit.
Moreover, the limiting odd series happens to terminate because of a
$\Gamma$ function in the denominator.

\begin{figure}[htbp]
$$
\begin{picture}(80,50)(-40,-25)
\put(0,0){\circle{40}}
\put(0,0){\oval(36,36)[b]}
\put(-19,1){\line(-1,0){10}}
\put(-19,-1){\line(-1,0){10}}
\put(19,1){\line(1,0){10}}
\put(19,-1){\line(1,0){10}}
\put(-19,0){\circle*4}
\put(19,0){\circle*4}
\put(-3,9)2
\put(-3,-29)4
\end{picture} $$
\caption{ \label{one-loop}
The one-loop diagram where the asymptotic on-shell
expansion in $m_2\ll m_4$, $p^2 = -m_4^2$, is performed.}
\end{figure}
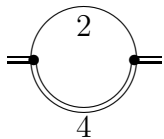

The use of eq.~(\ref{odd}) gives us an alternative way of achieving
the same result, without specifying the weight function and avoiding
the necessity to evaluate the indeterminacy that arises at
half-integer negative $n$ in eq.(\ref{even}) because of a singular
coefficient.  We have verified both procedures in yet another case
when the integration in eq.~(\ref{even}) can be performed
analytically in terms of $\Gamma$ functions. This is the scalar
one-loop diagram with one light mass and a line on mass shell
(fig.~\ref{one-loop}) \footnote{On preparing the present paper, we
became aware that a general method of constructing typically
Minkowskian and on-shell asymptotic expansions was proposed by V. A.
Smirnov \cite{1loop} and demonstrated by one-loop examples (including
(fig.~\ref{one-loop} )) and by a non-trivial two-loop master-integral
\cite{2loop}.}.  For odd expansion coefficients we obtain the same
results either by analytically continuing eq.(\ref{even}) or by
reducing the dimension of integration to $N-1$ via eq.~(\ref{odd}).
The complete expansion, which includes the contributions of subgraphs
\{4\} and \{2, 4\}, perfectly reproduces the exact integral in the
Feynman parameter

\begin{eqnarray}
&&(\mu^2)^\varepsilon \int \frac{d^N k }
{\pi^{N/2} \Gamma(3-N/2) (k^2 + m_2^2) (-2pk+k^2) }
\nonumber \\*
&& =
\frac{1}{\varepsilon} + 2 - \ln\left(\frac{-p^2}{\mu^2} \right)
- r^2\ln r - r^2\sqrt{4/r^2-1} \arctan\left(\sqrt{4/r^2-1}\right)
+ {\cal O}(\varepsilon)
\nonumber \\*
&&  =
\frac{1}{\varepsilon} + 2 - \ln\left(\frac{-p^2}{\mu^2} \right)
- \pi \left(r + \frac{r^3}{8} + \frac{r^5}{128} \right)
- r^2 \left( \ln r -1 \right) - \frac{r^4}{12} + ...,
\end{eqnarray}

\noindent
where $r^2 = m_2^2/(-p^2)$.
It is worth pointing out the fact that analytic continuation
recovers just the causal rule of getting around the singularity, as in
eq.~(\ref{odd}), which is automatically implied in the Euclidean space
where $k_2^2\ge 0$ after the Wick rotation.

Relations (\ref{c4})--(\ref{odd}) are sufficient [with a trivial
generalization of eq.~(\ref{c4}) by binomial coefficients as index
$j_4$ is arbitrary] for constructing the asymptotic expansion terms
for the ultraviolet subgraph \{4\} of the prototype,
fig.~\ref{loop}(b).  The procedure does not require any modifications
for a higher-loop analog of fig.~\ref{loop}(a) with an arbitrary
insertion in line~2.

The expansion of ultraviolet subgraph $\{ 4 \}$ suffers from on-shell
infrared singularities. Therefore, the expansion is not perfectly local in
coordinate space (or polynomial in momentum-space). Along with regular
Taylor-like terms, $(k_2^2)^n$, partially local on-shell infrared
counterterms $\delta^{(2n)}(pk_2)$ are produced in odd coefficients.
At $j_4 > 1$, also nonlocal negative powers of $k_2^2$ appear in
even coefficients. However, the general structure of the expansion as a
sum over ultraviolet subgraphs remains valid.

In case of the conventional dimensional regularization the sum over all
the ultraviolet subgraphs of the diagram (fig.~\ref{loop}) exactly
reproduces the expansion presented in ref. \cite{broad}.

\section{Techniques and results of the calculations}

The calculations in dimensional reduction do not generate any new
prototypes of Feynman diagrams. Just the coefficients before the same
scalar integrals are slightly changed for the bare diagrams. If we use
$N$-dimensional indices, the full contribution of the vectors and
$\varepsilon$ scalars to a diagram without external gluon lines
is obtained by replacing $N$ in the numerator of a scalarized expression
with 4. The contribution that involves $\varepsilon$ scalars can be
extracted by replacing $N^n$ with $4^n-(4-2\varepsilon)^n$.

If we need to separate the proper-vector and $\varepsilon$-scalar
propagators we should proceed as follows. The gluon propagator diagram is
scalarized by contracting it with the metric tensor either in the formal
four-dimensional space that contains the $(4-2\varepsilon)$-dimensional
vector and $\varepsilon$-scalar subspaces, or just in $4-2\varepsilon$
dimensions. According to the Ward--Slavnov--Taylor identities, the
vector part (for the sum of all diagrams in the relevant
order of perturbation theory) ought to be transverse with respect to the
external momentum, while the $\varepsilon$-scalar part is always
diagonal as there are no momenta in this subspace. Again, we perform the
algebra using the formal N-dimensional indices. As the external indices
are yet free, we should set $N=4$, in order to sum up the vector and
$\varepsilon$-scalar contributions in the internal lines of the diagram.
Then the external indices are contracted, which produces $N^n$ with
$n=0$,~1.  The $\varepsilon$-scalar propagator is then obtained by
replacing $N^n$ with $[4^n - (4-2 \varepsilon)^n]/(2\varepsilon)$.  The
scalar coefficient before the transverse projection operator in the
vector propagator can be evaluated by replacing $N^n$ with
$(4-2\varepsilon)^n/(3-2 \varepsilon)$. The momenta are always assumed
to belong to $4-2\varepsilon$ dimensions.

According to eq.~(\ref{2loop}), the pole mass acquires additive
contributions from the two-loop diagrams of the quark propagator.
The evaluation of the $\varepsilon$-scalar contribution to the
one-loop diagram and its derivative on shell is an easy task
performed analogously to the corresponding calculation in the
conventional dimensional regularization \cite{else,broad}. The
construction of the small mass asymptotic expansion for the diagram
of fig.~\ref{loop}(a) was discussed above. Other two-loop diagrams
shown in fig.~\ref{others} are, in principle,  computed
straightforwardly on the same lines as in the conventional
dimensional regularization \cite{broad}. The unrenormalized results
for individual bare two-loop diagrams are summarized in the Appendix.
For comparison we also present the corresponding contributions in the
conventional dimensional regularization.

Here are our results in the framework of the dimensional reduction
MMS scheme with the zero pole mass for $\varepsilon$ scalars. We have
evaluated the one-loop vertex and field renormalization constants:

\begin{eqnarray}
Z_{ffv} & = & 1 - \frac{h}{\varepsilon} \Big[
\Big(\frac{\xi-1}{4} + 1 \Big) C_a + \xi C_f \Big] + {\cal O} (h^2),
\label{Zffv}
\\
Z_{ffs} & = & 1 - \frac{h}{\varepsilon} \Big[
\frac{\xi-1}{2} C_a + \left(\xi+1\right) C_f \Big] + {\cal O} (h^2),
\label{Zffs}
\\
Z_v & = & 1 - \frac{h}{\varepsilon} \Big[
\Big( \frac{5}{3} - \frac{\xi-1}{2} \Big) C_a
- \frac{2}{3} N_f \Big] + {\cal O} (h^2),
\label{Zv}
\\
Z_s & = & 1 - \frac{h}{\varepsilon} \Big[
\left(3 - \xi \right) C_a - N_f \Big] + {\cal O} (h^2).
\label{Zs}
\end{eqnarray}

\noindent
Indices $f$, $v$, $s$ denote the type of the particles, fermions,
vectors, and $\varepsilon$ scalars. The expansion parameter is
$h=g^2/(16 \pi^2) = \alpha_s/(4 \pi)$; $\xi$ is the parameter of
the covariant gauge, $\xi=1$ corresponds to the Feynman gauge,
$\xi=0$ to the Landau gauge. We assume an arbitrary simple Lie gauge
group of colors and an arbitrary irreducible representation for the
quarks. In case of $SU(n)$ and the fundamental representation, the group
coefficients would be: $C_a = n$, $C_f = (n^2-1)/(2n)$.  The trace
normalization was chosen as $T_f  = 1/2$. The number of quark flavors
is $N_f$. The field renormalization constants are defined for the
one-par\-ticle-ir\-re\-duc\-ible two-point diagrams, including one leg.
They enter the charge renormalization in positive powers:

\begin{eqnarray}
Z_h & = & Z_{ffv}^2 Z_f^2 Z_v ~=~
1 + \frac{h}{\varepsilon} \Big(
- \frac{11}{3} C_a + \frac{2}{3} N_f \Big) + {\cal O} (h^2),
\label{Zh} \\
Z_Y & = & Z_{ffs}^2 Z_f^2 Z_s ~=~
1 + \frac{h}{\varepsilon} \Big(
- 2 C_a - 2 C_f + N_f \Big) + {\cal O} (h^2).
\label{ZY}
\end{eqnarray}

\noindent As expected, the gauge and Yukawa charges are renormalized
differently, eqs.~(\ref{Zh}) and (\ref{ZY}), because there is no
supersymmetry to protect the tree-level coincidence of the
$\varepsilon$-scalar and vector coupling constants. We everywhere replace
the Yukawa charge with the corresponding function of the gauge charge,
that is, exploit a special solution of the renormalization-group equations.

The nonminimal mass counterterm for the $\varepsilon$ scalars

\begin{equation}
\Delta m^2_s = - h \sum_{f=1}^{N_f}
m_f^2 \left(\frac{2}{\varepsilon}  + 2  - 2 \ln \frac{m_f^2}{\mu^2}
\right) + {\cal O} (\varepsilon) + {\cal O} (h^2)
\label{ms}
\end{equation}

\noindent is necessary to insure that their pole mass is zero.

The fermion field and mass renormalization constants were calculated
up to two loops:

\begin{eqnarray}
Z_f & = & 1 - \frac{h}{\varepsilon} \Big( - \xi C_f \Big)
\nonumber \\*[2mm]
& - &h^2 \Big\{  \Big[ \frac{1}{\varepsilon^2} - \frac{13}{4 \varepsilon}
+ (\xi-1) \Big( \frac{5}{4\varepsilon^2} - \frac{5}{4 \varepsilon} \Big)
+ (\xi-1)^2 \Big( \frac{1}{4\varepsilon^2} - \frac{1}{8 \varepsilon} \Big)
\Big] C_a C_f
\nonumber \\*[2mm]
&& +
\Big[
\frac{1}{2 \varepsilon^2} + \frac{7}{4 \varepsilon}
+ \frac{\xi-1}{\varepsilon^2} + \frac{(\xi-1)^2}{2\varepsilon^2}
\Big] C_f^2 \Big\} + {\cal O} (h^3),
\label{Zf}
\\ [2mm]
Z_m & = & 1 - \frac{h}{\varepsilon} 3 C_f
- h^2 \Big[
\Big( -\frac{11}{2\varepsilon^2} + \frac{79}{12 \varepsilon} \Big) C_a C_f
\nonumber \\* [2mm]
&&
+
\Big( -\frac{9}{2\varepsilon^2} - \frac{1}{4 \varepsilon} \Big) C_f^2
+
\Big( \frac{1}{\varepsilon^2} - \frac{1}{3 \varepsilon} \Big) C_f N_f
\Big] + {\cal O} (h^3).
\label{Zm}
\end{eqnarray}

\noindent
In spite of the fact that there are several masses in the theory, there
exist a common mass renormalization constant for all flavors. Let us
remind that in the conventional dimensional regularization the mixing of
the masses through the diagram of fig.~\ref{loop}(a) was prevented by
the gauge invariance which forbids a mass term in the vector
propagator.  In case of the $\varepsilon$ scalars, there is no such
ban in general. But subtracting the $\varepsilon$-scalar propagator
at zero, we, at the same time, get rid of the $m^2/\varepsilon$ term
in the incomplete $R$ operation. This is related to the fact that the
subtraction at zero insures the absence of an infrared singularity in
the renormalized subgraph $\{1, 2, 5\}$.

The final result for the pole mass of a quark with a given flavor $f$
is

\begin{eqnarray}
\Big( \frac{m_P}{m} \Big)_f & = & 1
+ h \Big[ 5 - 3 \ln \frac{m_f^2}{\mu^2}
\Big] C_f
\nonumber \\*[2mm]
& + & h^2 \Big\{
\Big[ \frac{1093}{24} + \frac{11}{2}\ln^2 \frac{m_f^2}{\mu^2}
- \frac{179}{6}\ln \frac{m_f^2}{\mu^2}
\nonumber \\*[2mm]
&-& 6 \zeta(3) - 8 \zeta(2) + 24 \zeta(2) \ln 2 \Big] C_a C_f
\nonumber \\*[2mm]
& + &
\Big[ -\frac{59}{8}
+ \frac{9}{2}\ln^2  \frac{m_f^2}{\mu^2}
+ \frac{3}{2}\ln \frac{m_f^2}{\mu^2}
\nonumber \\*[2mm]
& + &
12 \zeta(3) + 30 \zeta(2) - 48 \zeta(2) \ln 2 \Big] C_f^2
\nonumber \\*[2mm]
& + & \sum_{j = 1}^{N_f} \Big[
- \ln^2 \frac{m_f^2}{\mu^2}
+ \frac{13}{3} \ln \frac{m_f^2}{\mu^2}
+ E(m_j/m_f) \Big] C_f \Big\} + {\cal O} (h^3).
\label{mP}
\end{eqnarray}

\noindent
The sum over flavors represents the renormalized contribution of the
diagram with the quark loop, fig.~\ref{loop}(a). We know this
contribution in three kinds of expansions. The small-mass expansion,
the construction and meaning of which was discussed above, gives

\begin{eqnarray}
E(r) & = & - \frac{37}{6} - 4 \zeta (2) + 2 \pi^2 r - 12 r^2 + 2 \pi^2 r^3
\nonumber \\*[3mm]
&& + \Big[ -\frac{151}{18} - 4 \zeta(2) + \frac{13}{3} \ln r^2 - \ln^2 r^2
\Big] r^4
+ \Big[ \frac{76}{75} - \frac{8}{15} \ln r^2 \Big] r^6
\nonumber \\*[3mm]
&&
+ \Big[ \frac{1389}{9800} - \frac{9}{70} \ln r^2 \Big] r^8
+ \Big[ \frac{3988}{99225} - \frac{16}{315} \ln r^2 \Big] r^{10}
\nonumber \\*[3mm]
& + & {\cal O} (r^{12} \ln r^2).
\label{small}
\end{eqnarray}

\noindent
Numerically the most essential correction due to the mass of a light
quark corresponds to the linear term in the mass ratio.

The large-mass expansion, performed according to the Euclidean rules,
yields

\begin{eqnarray}
E(r) & = & \frac{20}{9} + \frac{13}{3} \ln r^2 + \ln^2 r^2
- \Big( \frac{76}{75} + \frac{8}{15} \ln r^2 \Big)/r^2
\nonumber \\*[2mm]
&&
- \Big( \frac{1389}{9800}  + \frac{9}{70} \ln r^2 \Big)/r^4
- \Big( \frac{3988}{99225}  + \frac{16}{315} \ln r^2 \Big)/r^6
\nonumber \\*[2mm]
&&
- \Big( \frac{1229}{78408}  + \frac{5}{198} \ln r^2 \Big)/r^8
- \Big( \frac{184452}{25050025}  + \frac{72}{5005} \ln r^2 \Big)/r^{10}
\nonumber \\*[2mm]
&&
+ {\cal O} (r^{-12} \ln r^2).
\label{large}
\end{eqnarray}

\noindent
In most calculations, people simply ignore the quarks that
are heavier than the characteristic energy scale of the problem. They
use thus an effective low-energy theory with the reduced number of
particles. The running charges and masses of this effective theory are
related to those of the full theory by the so-called matching
conditions that the running parameters are supposed to change their
running rate at some value of the renormalization scale $\mu$ of the
order of the heavy-particle mass. Of course, the low-energy
effective theory is no longer valid at that scale. The matching
condition just ensures the equivalence of the two theories at low
energies where the heavy particles ought to decouple, up to power
corrections, irreproducible in the low-energy theory. The coincidence of
the pole mass in the two theories provides the two-loop matching condition
for the running mass. The charge matching is done by means of another
physical quantity, like the invariant charge. In the MMS scheme
(as well as $\overline{\rm MS}$), on the one-loop level, relevant here,
it is done at $\mu=m_j$ \cite{SM}. One should bear in mind, however,
implicit numeric ambiguities of any matching condition, similar to the
scheme dependence, owing to the fact that higher orders of perturbation
theory are unavailable. The result, thus, depends on the choice of a physical
quantity and on the detailed prescription of equating it at low energy.
When defining the two-loop mass matching, the term of order zero in the
ratio of the masses is absorbed into the matching point. The actual
correction starts from the leading power $m_f^2/m_j^2$.

\begin{figure}[htbp]
\centerline{\vbox{\epsfysize=4.5in
\epsfbox{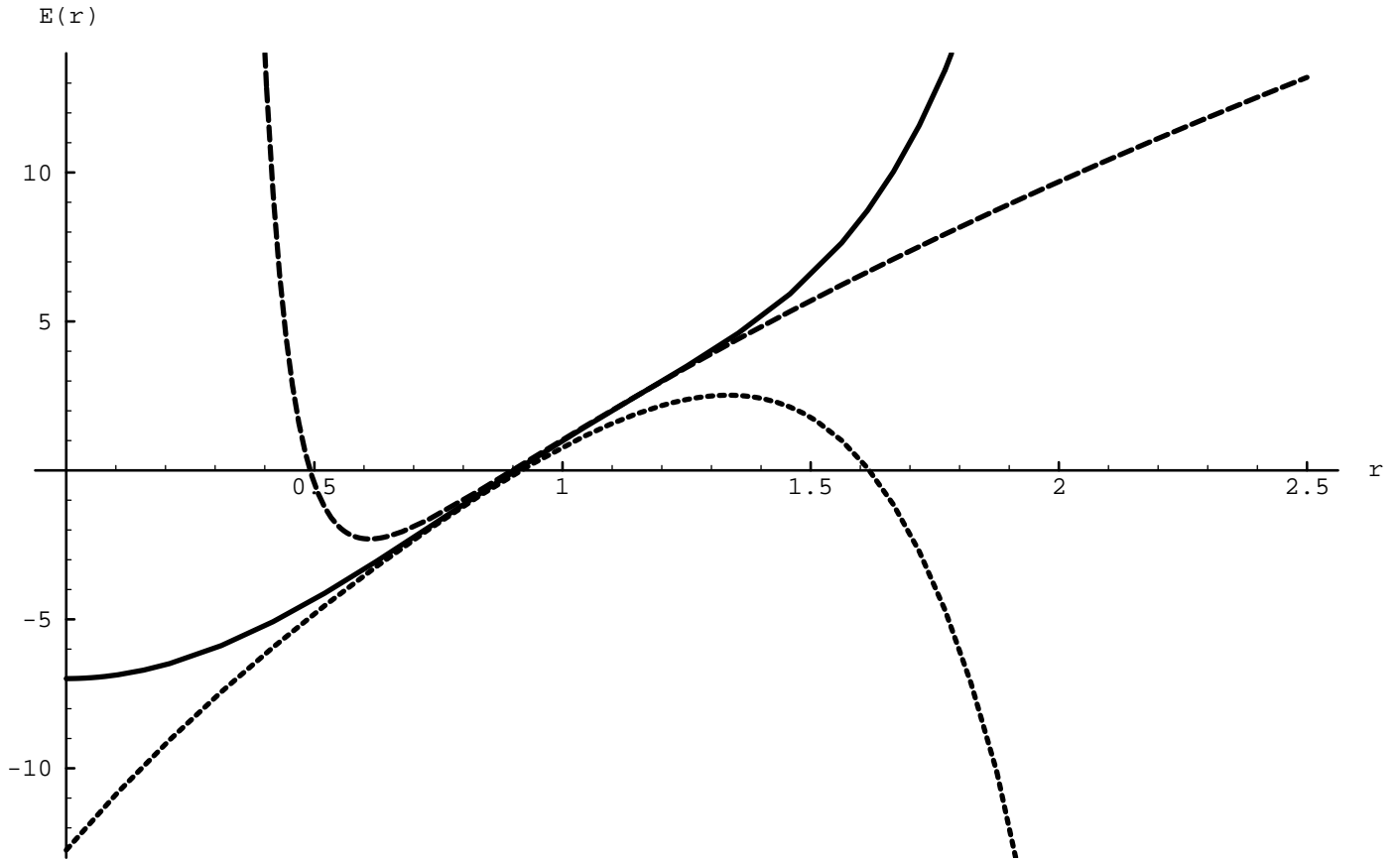}
}}
\caption{\label{expansions}
Three expansions for the renormalized diagram of fig.~\ref{loop}(a)
in the dimensional reduction MMS scheme. The dotted line corresponds to
the small-mass expansion up to $r^6$, the dashed line to the large-mass
expansion up to $1/r^6$, the solid line to the intermediate expansion
up to $(r^2-1)^3$.}
\end{figure}

The third available expansion is performed in the difference of the
squared masses. Such an expansion covers the intermediate region
between eqs.~(\ref{small}) and (\ref{large}). In the present case the
intermediate expansion is just a regular Taylor expansion involving
only the whole diagram as the ultraviolet subgraph. A surprising remark
is, however, that generating this trivial expansion is the most
time-consuming of the three, while the most complicated small-mass
expansion is the fastest.

\begin{eqnarray}
E(r) & = &  - \frac{73}{6} + 8 \zeta(2)
+ \Big[ - 8 + 8 \zeta(2) \Big] (r^2-1)
- \Big[ 1 + \frac{1}{2} \zeta(2) \Big] (r^2-1)^2
\nonumber \\*[2mm]
& + & [r^2-1]^3
- \Big[ \frac{1}{2} + \frac{3}{32} \zeta(2) \Big] (r^2-1)^4
+ \Big[ \frac{19}{60} + \frac{3}{32} \zeta(2) \Big] (r^2-1)^5
\nonumber \\*[2mm]
& + & {\cal O} [(r^2-1)^6].
\label{medium}
\end{eqnarray}

\noindent
As the masses of different quarks are quite different, higher terms of
eq.~(\ref{medium}) are of little practical use. The value that is
important is $E(1)$ which defines the self-contribution of a quark to
its pole mass. The intermediate expansion just demonstrates the means of
recovering the function $E(r)$ in the whole region of the mass ratio.
The matching of the three expansions is demonstrated in
fig.~\ref{expansions}.  We terminated the expansions at the same
relatively low order for the sake of clearness.  Keeping more terms,
we get a wider region around $r=1$ where the curves practically
coincide.

\section{Discussion}

The results of the three expansions (\ref{small})--(\ref{medium})
differ from the corresponding expansions in the conventional
dimensional regularization MMS scheme only by a constant
$-\frac{1}{4}$ in the zeroth order, which confirms that these are
just two different renormalization schemes.

All the three expansions reproduce the same dependence of the diagram
on $\mu$. This confirms the fact that our renormalization scheme,
though being nonminimal, remains massless. The latter implies that the
renormalization-group functions do not depend on masses and other
dimensional coupling constants but polynomially. The nonminimality
of our scheme nevertheless reveals itself by the fact that the
$\beta$ function of the fermion mass cannot be extracted from the
first-order pole of the renormalization constant (\ref{Zm}). The
correct way of defining the mass beta function (valid in an arbitrary
renormalization scheme) is to extract it from the condition of the
renormalization invariance of a physical quantity, the pole mass,
by differentiating eq.~(\ref{mP}) with respect to $\ln\mu^2$. The result

\begin{equation}
\beta_m = m \Big[ h \Big(  - 3 C_f \Big)
+ h ^2 \Big(  - \frac{23}{2} C_a C_f - \frac{3}{2} C_f^2 + C_f N_f \Big)
+ {\cal O} (h^3) \Big]
\label{beta_m}
\end{equation}

\noindent
does not coincide with  what has been obtained in ref.~\cite{beta_m}
in the minimal subtraction scheme, although the renormalization
constant (\ref{Zm}), the one-loop counterterms,
(\ref{Zffv})--(\ref{ZY}), and the one-loop contribution to
eq.~(\ref{Zf}), agree with refs.~\cite{beta_m,beta1}, which confirms
that all the calculations are correct.

The situation is explained quite evidently indeed. In our
renormalization scheme the bare mass of the fermion ceases to be a
renormalization-invariant quantity. The point is that the physical
mass involves contributions from two mutually correlated bare
masses -- of the quark and of the $\varepsilon$ scalars. Both {\em do}
depend on $\mu$, to ensure that the physical quantity is invariant.
The true reason for the difference between the $\beta$ function and the
$1 /\varepsilon$ pole in the renormalization constant
is thus an extra bare parameter as compared to the
number of the physical renormalized parameters. In fact, the
renormalized mass of the $\varepsilon$ scalars becomes one more
independent quantity. However, we keep it equal to zero, while in the
minimal subtraction scheme it would be running as well.
Total subtraction of loop corrections to the $\varepsilon$-scalar mass,
as we perform here, seems to be practically the only scheme applicable
to physical calculations (including the finite parts) with masses in
nonsupersymmetric theories, if we are going to stick as closely as
possible to the original idea of dimensional reduction which naturally
suggests the zero tree-level mass for the $\varepsilon$ scalars as well
as for the gauge vectors. The minimal subtractions, being admissible
theoretically and convenient for re\-norm\-al\-iz\-a\-tion-group
calculations, would involve an extra independent mass parameter,
irrelevant to actual physics, however, up to a scheme redefinition of
other parameters \cite{jones}.

In principle, all renormalization schemes are, of course, equivalent.
The difference of our $\beta$ function (\ref{beta_m}) from
ref. ~\cite{beta_m} is just a re\-norm\-al\-iz\-a\-tion-scheme
difference which can be compensated for by a finite recalculation
of the parameters.

Our renormalization scheme can be related by proper
recalculation to the standard dimensional renormalization (without the
$\varepsilon$ scalars) as well.
A recalculation can always be done by equating a physical quantity
evaluated in the two schemes. Our result (\ref{mP}) allows us to
recalculate the renormalized mass at the two-loop level. However, we
have to take into account the fact that the renormalized charges
in the two schemes are also different. To relate them we  should
equate another renormalization-invariant quantity like the one-loop
invariant charge. The necessary relation between the renormalized charge
in dimensional reduction and in the conventional dimensional
renormalization can be taken from ref.~\cite{rho}:

\begin{equation}
h_{\rm DRED} = \Big\{
h \Big[ 1 + \frac{1}{3} C_ah  + {\cal O} (h^2) \Big]
\Big\}_{\rm DREG}.
\end{equation}

\noindent
Then the mass recalculation looks like

\begin{equation}
m_{\rm DRED} =
\Big\{ m \Big[ 1 - h C_f
+  h^2 C_f \Big( -\frac{11}{12} C_a - \frac{5}{2} C_f
+ \frac{1}{4} N_f \Big)  + {\cal O}(h^3)  \Big] \Big\}_{\rm DREG} .
\label{recalculation}
\end{equation}

One could also derive the recalculation formulae from comparing
the re\-norm\-al\-iz\-a\-tion-group functions in the different schemes.
However, then one would need to calculate one-loop further. For
example, the comparison of the two-loop $\beta$ function (\ref{beta_m})
with the corresponding result in the conventional dimensional
renormalization reproduces the one-loop term of eq.~(\ref{recalculation}).

{\bf Acknowledgments}
~~~This paper was supported in part by RFFI grant \# 96-02-17531,
JSPS FSU Project and the Heisenberg-Landau program,
and on the part of Kal\-my\-kov, by Volkswagenstiftung.
Kalmykov is grateful to the Physics Department of Bielefeld University
for warm hospitality.

\setcounter{section}{0}
\setcounter{equation}{0}
\renewcommand{\theequation}{A.\arabic{equation}}
\renewcommand{\thesection}{APPENDIX A. }
\section{Contributions of bare two-loop diagrams to the pole mass of a quark}

We present the unrenormalized contributions, implying a common factor
$h^2 m (\mu^2/m^2)^{2 \varepsilon}$,  up to discarded ${\cal O} (\varepsilon)$
terms, in the regularization by dimensional reduction and the
conventional dimensional regularization. In the former case we use
the parentheses, while in the latter case the brackets, in the notation
of the diagrams.

\begin{eqnarray}
\label{rdr1a}
{\rm fig.~\ref{loop}(a)}
& = & C_f
\Big[ - \frac{1}{\varepsilon^2} - \frac{4}{\varepsilon}
- \frac{19}{3}
+ \Big( \frac{m_1^2}{m^2} \Big)^{1-\varepsilon}
\Big( \frac{2}{\varepsilon} + 6 \Big) +  E(m_1/m)  \Big ],
\\*[2mm]
{\rm fig.~\ref{loop}[a] }
& = & C_f
\Big[ -\frac{1}{\varepsilon^2} - \frac{7}{2 \varepsilon}
- \frac{61}{12}  + E(m_1/m)  \Big ],
\end{eqnarray}

\noindent
where $m_1$ is the mass of the quark in the loop, and $E(r)$ is defined
by expansions (\ref{small})--(\ref{medium}). Note that the result in
dimensional reduction (\ref{rdr1a}) involves a mass-mixing term
proportional to $m_1^2$. This term is, however, canceled latter by the
$\varepsilon$-scalar mass counterterm. Also, the $\varepsilon$-scalar
field renormalization counterterm affects the $1/\varepsilon$ pole of
the renormalized diagram, so that in the end the difference from the
conventional dimensional renormalization is reduced to $-\frac{1}{4}$.
The subsequent diagrams in the two schemes similarly involve differences
in the lower-order poles and the finite parts:

\begin{eqnarray}
{\rm fig.~\ref{others}(a+b)}
& = & C_a C_f \Big\{
(\xi - 1)^2 \Big( \frac{3}{8 \varepsilon} + \frac{23}{16} \Big)
\nonumber \\*[2mm]
& + &
(\xi - 1) \Big[- \frac{3}{4 \varepsilon^2} - \frac{3}{8 \varepsilon}
+ \frac{27}{16} - 3 \zeta(2)  \Big]
\nonumber \\*[2mm]
& + &
\frac{5}{2 \varepsilon^2} + \frac{41}{4 \varepsilon}
+ \frac{263}{8} + 10 \zeta(2)
 \Big\},
\\[2mm]
{\rm fig.~\ref{others}[a+b] }
& = & C_a C_f \Big\{
(\xi - 1)^2 \Big( \frac{3}{8 \varepsilon} + \frac{23}{16} \Big)
\nonumber \\*[2mm]
& + &
(\xi - 1) \Big[- \frac{3}{4 \varepsilon^2} + \frac{1}{8 \varepsilon}
+ \frac{31}{16} - 3 \zeta(2)  \Big]
\nonumber \\*[2mm]
& + &
\frac{5}{2 \varepsilon^2} + \frac{39}{4 \varepsilon}
+ \frac{261}{8} + 10 \zeta(2)
 \Big\};
\\[2mm]
{\rm fig.~\ref{others}(c) }
 & = & C_f^2 \Big\{
(\xi - 1)^2 \Big(\frac{3}{4 \varepsilon} + \frac{23}{8} \Big)
\nonumber \\*[2mm]
& + &
(\xi - 1) \Big[- \frac{3}{\varepsilon^2} - \frac{13}{2 \varepsilon}
- \frac{63}{4} - 12 \zeta(2)  \Big]
\nonumber \\*[2mm]
& + &
\frac{21}{2 \varepsilon^2} + \frac{85}{4 \varepsilon}
+ \frac{459}{8} + 6 \zeta(2)
 \Big\},
\\[2mm]
{\rm fig.~\ref{others}[c] }
& = & C_f^2 \Big\{
(\xi - 1)^2 \Big( \frac{3}{4 \varepsilon} + \frac{23}{8} \Big)
\nonumber \\*[2mm]
& + &
(\xi - 1) \Big[- \frac{3}{\varepsilon^2} - \frac{11}{2 \varepsilon}
- \frac{61}{4} - 12 \zeta(2) \Big]
\nonumber \\*[2mm]
& + &
\frac{21}{2 \varepsilon^2} + \frac{65}{4 \varepsilon}
+ \frac{443}{8} + 6 \zeta(2)
 \Big\};
\\[2mm]
{\rm fig.~\ref{others}(d)}
& = & C_a C_f
\Big\{
(\xi - 1)^2 \Big( -\frac{3}{4 \varepsilon} - \frac{23}{8} \Big)
\nonumber \\*[2mm]
& + &
(\xi - 1) \Big[\frac{9}{4 \varepsilon^2} + \frac{29}{8 \varepsilon}
+ \frac{99}{16} +9 \zeta(2) \Big]
\nonumber \\*[2mm]
& + &
\frac{9}{2 \varepsilon^2} + \frac{67}{4 \varepsilon}
+ \frac{405}{8} - 6 \zeta(2)
 \Big\},
\\[2mm]
{\rm fig.~\ref{others}[d] }
& = & C_a C_f
\Big\{
(\xi - 1)^2 \Big( -\frac{3}{4 \varepsilon} - \frac{23}{8} \Big)
\nonumber \\*[2mm]
& + &
(\xi - 1) \Big[\frac{9}{4 \varepsilon^2} + \frac{21}{8 \varepsilon}
+ \frac{91}{16} + 9 \zeta(2) \Big]
\nonumber \\*[2mm]
& + &
\frac{9}{2 \varepsilon^2} + \frac{57}{4 \varepsilon}
+ \frac{343}{8} - 6 \zeta(2)
 \Big\};
\\[2mm]
{\rm fig.~\ref{others}(e) }
& = & C_f \Big ( C_f - \frac{1}{2} C_a  \Big )
\Big\{
(\xi - 1)^2 \Big( - \frac{3}{4 \varepsilon} - \frac{23}{8} \Big)
\nonumber \\*[2mm]
& + &
(\xi - 1) \Big[\frac{3}{\varepsilon^2} + \frac{13}{2 \varepsilon}
+ \frac{63}{4} + 12 \zeta(2) \Big]
+ \frac{3}{\varepsilon^2} + \frac{15}{2 \varepsilon} + \frac{53}{4}
\nonumber \\*[2mm]
& + & 24 \zeta(2) - 48 \zeta(2) \ln 2 + 12 \zeta(3)
 \Big\},
\\[2mm]
{\rm fig.~\ref{others}[e] }
& = & C_f \Big ( C_f - \frac{1}{2} C_a  \Big )
\Big\{
(\xi - 1)^2 \Big( - \frac{3}{4 \varepsilon} - \frac{23}{8} \Big)
\nonumber \\*[2mm]
& + &
(\xi - 1) \Big[\frac{3}{\varepsilon^2} + \frac{11}{2 \varepsilon}
+ \frac{61}{4} + 12 \zeta(2) \Big]
+ \frac{3}{\varepsilon^2} + \frac{5}{2 \varepsilon}
- \frac{1}{4}
\nonumber \\*[2mm]
& + & 24 \zeta(2) - 48 \zeta(2) \ln 2 + 12 \zeta(3)
 \Big\}.
\end{eqnarray}


\begin{thebibliography}{99}
\bibitem{Siegel}
W. Siegel, Phys. Lett. B {\bf 84} (1979) 193.

\bibitem{inconsistency}
W. Siegel, Phys. Lett. B {\bf 94} (1980) 37.

\bibitem{AV}
L. V. Avdeev and A. A. Vladimirov, Nucl. Phys. B {\bf 219} (1983), 262;

L. Avdeev, Phys. Lett. B {\bf 117} (1982), 317.

\bibitem{Fierz}
L. V. Avdeev, Theor. Math. Phys. {\bf 58} (1984) 203.

\bibitem{Italy}
G. Altarelli, G. Curci, G. Martinelli and S. Petrarca,
Nucl. Phys. B {\bf 187} (1981) 461.

\bibitem{rho}
L. Avdeev, J. Fleischer, S. Mikhailov and O. Tarasov,
Phys. Lett. B {\bf 336} (1994) 560;
Erratum B {\bf 349} (1995) 597.

\bibitem{else}
R. Tarrach, Nucl. Phys. B{\bf 183} (1981) 384.

\bibitem{broad}
N. Gray, D. J. Broadhurst, W. Grafe and K. Schilcher,
Z. Phys. C - Particles and Fields {\bf 48} (1990) 673.

\bibitem{rules}
F. V. Tkachov,
Institute for Nuclear Research preprint, P-0332,
``Euclidean Asymptotics of Feynman integrals: Basic notations.'',
Moscow (1983);
P-0358,
``Asymptotics of Euclidean Feynman integrals. 2. one loop case.'',
Moscow (1984);
Int. J. Mod. Phys. A8 (1993), 2047;

K. G. Chetyrkin and V. A. Smirnov,
Institute for Nuclear Research preprint, P-0518,
``Asymptotic expansions of Feynman amplitudes, $R^\ast$-operation and
the method of glueing'', (in Russian),  Moscow (1987);

S. G. Gorishny and S. A. Larin,
Nucl. Phys. B283 (1987), 452;

K. G. Chetyrkin,
Teor. Math. Phys. 75 (1988), 26; ibid 76 (1988), 207;
Max-Planck-Institute preprint, MPI-PAE/PTh-13/91,
``Combinatorics of $R-, R^{-1}-$, and $R^\ast-$ operations and
asymptotic expansions of Feynman integrals in the limit of large
momenta and masses'', Munich (1991);

V. A. Smirnov,
Mod. Phys. Lett. A 3 (1988), 381;
Comm. Math. Phys. 134 (1990), 109;
{\it Renormalization and asymptotic expansions}
(Bikrh\"auser, Basel, 1991);

G. B. Pivovarov and F. V. Tkachov,
Int. J. Mod. Phys. A8 (1993), 2241.

\bibitem{jones}
I. Jack and D. R. T. Jones,
Phys. Lett. B {\bf 333} (1994) 372;

I. Jack, D. R. T. Jones, S. P. Martin, M. T. Vaughn and Y. Yamada,
Phys. Rev. D {\bf 50} (1994) 5481.

\bibitem{ms}
D. J. Broadhurst,
Z. Phys. C - Particles and Fields {\bf 54} (1992) 599.

\bibitem{shell2}
J. Fleischer and O. V. Tarasov,
Comp. Phys. Commun. 71 (1992) 193.

\bibitem{proto}
L. V. Avdeev, Com. Phys. Commun. {\bf 98} (1996) 15.

\bibitem{parts}
K. G. Chetyrkin and F. V. Tkachov,
Nucl. Phys. B {\bf 192} (1981) 159;

F. V. Tkachov,
Phys. Lett. B {\bf 100} (1981) 65.

\bibitem{1loop}
V. A. Smirnov,  Phys. Lett. B {\bf 394} (1997) 205.

\bibitem{2loop}
A. Czarnecki and V. A. Smirnov, Phys. Lett. B {\bf 394} (1997) 211.

\bibitem{SM}
D. V. Shirkov and S. V. Mikhailov,
Z. Phys. C - Particles and Fields {\bf 63} (1994) 463.

\bibitem{beta_m}
I. Jack, D. R. T. Jones and K. L. Roberts,
Z. Phys. C - Particles and Fields {\bf 63} (1994) 151.

\bibitem{beta1}
I. Jack, D. R. T. Jones and K. L. Roberts,
Z. Phys. C - Particles and Fields {\bf 62} (1994) 161.

\end{thebibliography}
\end{document}